\documentclass[conference]{IEEEtran}
\IEEEoverridecommandlockouts
\usepackage{cite}
\usepackage{amsmath,amssymb,amsfonts}
\usepackage{algorithmic}
\usepackage{graphicx}
\usepackage{textcomp}
\usepackage{xcolor}
\usepackage[svgnames]{xcolor}
\usepackage{soul} 
\usepackage{tikz}
\usetikzlibrary{arrows.meta, positioning, calc, fit, backgrounds}
\usepackage{amsmath}
\def\BibTeX{{\rm B\kern-.05em{\sc i\kern-.025em b}\kern-.08em
		T\kern-.1667em\lower.7ex\hbox{E}\kern-.125emX}}
\definecolor{blue1}{RGB}{173, 216, 230}
\definecolor{orange1}{RGB}{255, 165, 0}
\definecolor{green1}{RGB}{144, 238, 144}
\definecolor{purple1}{RGB}{216, 191, 216}

\begin{document}
	
	\title{ A Synchronous EEG-fNIRS BCI: A Proof-of-Concept for Multimodal Avalanche Analysis of Motor Cognition in Older Adults
		\thanks{This work was supported by STI 2030 Major Projects and the National Key Research and Development Program of China (Grant No. 2022ZD0208500). Ethical approval was granted by Westlake University (Ethics No. 20191023Swan001) and Hangzhou Fist People's Hospital (Ethics No. 2025ZN244-1).}
	}
	
	\author{Eva Guttmann-Flury$^{1,2}$, Yun-Hsuan Chen$^2$, Qiaoyuan Xiang$^3$, Hao Zhang$^{3}$, Mohamad Sawan$^2$\\
		\textit{$^1$School of Integrated Circuits, Shanghai Jiao Tong University, Shanghai, China} \\
		\textit{$^2$CenBRAIN Neurotech, School of Engineering, Westlake University} \\
		\textit{$^3$Department of Neurology, Affiliated Hangzhou First People’s Hospital, Westlake University, School of Medicine} \\
		\textit{*Email: eva.guttmann.flury@gmail.com} \\

		}
		
		\maketitle
		
		\begin{abstract}
			This proof-of-concept study introduces a novel multimodal framework combining synchronized EEG-fNIRS modalities with neuronal avalanche analysis to identify early network dysfunction in Alzheimer's disease. The approach leverages complementary neural signals to examine motor network dynamics during execution and imagery tasks within an interactive task environment. Preliminary analysis of a small pilot cohort (N=4 subjects, including one with Mild Cognitive Impairment) validated the technical feasibility of the multimodal framework and revealed observable condition-dependent patterns in network organization. Two primary observations emerged: a reduced neural contrast between motor execution and imagery states, and increased trial-to-trial variability in network organization in the MCI participant. These initial results successfully validate the technical pipeline and provide hypothesis-generating observations for future statistically powered studies. The convergence of findings across modalities suggests that multimodal assessment of network flexibility may help detect functional changes in early Alzheimer's continuum, supporting the future development of this BCI-inspired framework into an engaging diagnostic tool.
			
		\end{abstract}
		
		
		\section{Introduction}
		The imperative for early, non-invasive biomarkers for Alzheimer's Disease (AD) is unmet by current diagnostic paradigms, which often identify the pathology only after significant, irreversible neurodegeneration. The prodromal stage, Mild Cognitive Impairment (MCI), presents a critical window for intervention, yet its heterogeneous progression and the high cost/low accessibility of gold-standard neuroimaging (e.g., amyloid-PET, fMRI) limit widespread screening. There is a clear need for cost-effective, sensitive tools that can probe the earliest signs of network-level brain dysfunction \cite{b1, b2, b3}.
		
		A compelling body of evidence now positions AD as a ``disconnection syndrome," characterized by the progressive failure of large-scale neural communication. This has motivated the search for functional biomarkers that can probe the integrity of these dynamic networks. The healthy brain is theorized to operate at a critical point, a dynamical regime poised between order and chaos that optimizes information processing and transmission. A key signature of this critical state is the presence of ``neuronal avalanches" -- spontaneous, aperiodic bursts of neural activity that propagate through the brain in cascades of all sizes, following a power-law distribution \cite{b4}.
		
		Critically, computational and empirical studies indicate that neurodegenerative pathologies, including AD, can disrupt this delicate critical balance. We therefore hypothesize that the precise spatiotemporal patterns of these neuronal avalanches carry vital information about the brain's functional health. The Avalanche Transition Matrix (ATM), which quantifies the probability of an avalanche spreading from one brain region to another, provides a novel and direct metric to track these large-scale communication pathways \cite{b5, b6, b7}.
		
		This proof-of-concept study therefore proposes that the properties of neuronal avalanches, derived from non-invasive electrophysiological recordings (EEG-fNIRS), can serve as sensitive biomarkers for early network dysfunction in AD. Building on evidence that AD disrupts critical brain dynamics, the study aims to test whether the breakdown of this critical state, measured through altered avalanche propagation and quantified by ATMs, can distinguish individuals with early cognitive impairment from healthy controls. This approach offers a promising new avenue for accessible, early detection.
		
		\section{Methods}
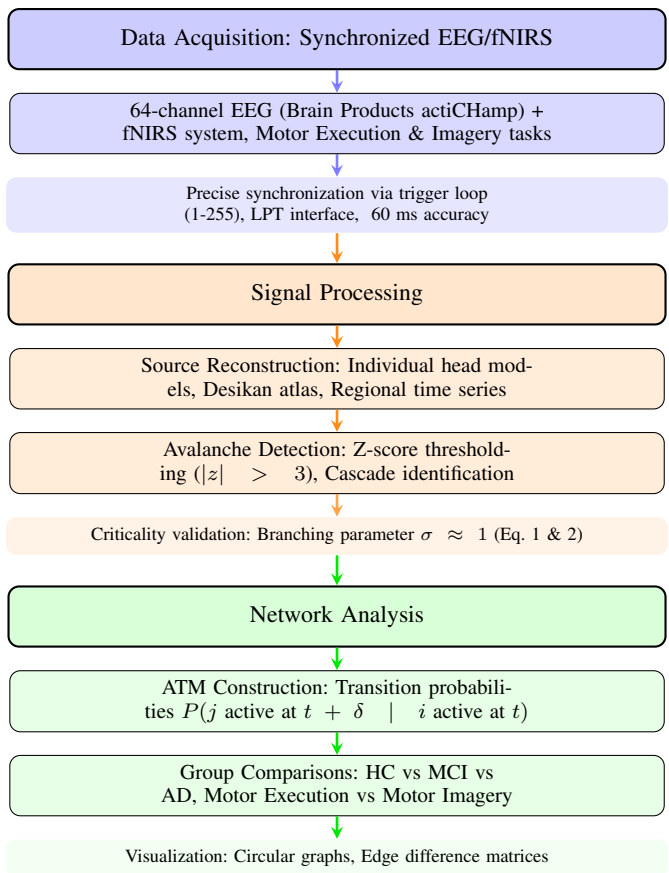
\begin{figure}[h!]
			\centering
			\begin{tikzpicture}[
				node distance=0.5cm and 0.3cm,
				title/.style={font=\bfseries\normalsize, align=center},
				section/.style={rectangle, rounded corners, minimum width=8.5cm, minimum height=0.8cm, text width=8.5cm, align=center, draw=black, thick, font=\small},
				process/.style={rectangle, rounded corners, minimum width=8.7cm, minimum height=0.7cm, text width=8.3cm, align=center, draw=black, thin, font=\footnotesize},
				arrow/.style={-stealth, thick, line width=1pt},
				detail/.style={rectangle, rounded corners, minimum width=8.8cm, minimum height=0.5cm, text width=7.8cm, align=center, font=\scriptsize}
				]
				
				\node[section, fill=blue!20] (acquisition) 
				{Data Acquisition: Synchronized EEG/fNIRS};
				\node[process, fill=blue!15, below=0.3 of acquisition] (devices) 
				{64-channel EEG (Brain Products actiCHamp) + fNIRS system, Motor Execution \& Imagery tasks};
				\node[detail, fill=blue!10, below=0.3 of devices] (sync) 
				{Precise synchronization via trigger loop (1-255), LPT interface, ~60 ms accuracy};
				
				\node[section, fill=orange!20, below=0.4 of sync] (preprocessing) 
				{Signal Processing};
				\node[process, fill=orange!15, below=0.3 of preprocessing] (source) 
				{Source Reconstruction: Individual head models, Desikan atlas, Regional time series};
				\node[process, fill=orange!15, below=0.3 of source] (avalanche) 
				{Avalanche Detection: Z-score thresholding ($|z| > 3$), Cascade identification};
				\node[detail, fill=orange!10, below=0.3 of avalanche] (criticality) 
				{Criticality validation: Branching parameter $\sigma \approx 1$ (Eq. 1 \& 2)};
				
				\node[section, fill=green!15, below=0.4 of criticality] (atm) 
				{Network Analysis};
				\node[process, fill=green!10, below=0.3 of atm] (matrix) 
				{ATM Construction: Transition probabilities $P(j\text{ active at }t+\delta \mid i\text{ active at }t)$};
				\node[process, fill=green!10, below=0.3 of matrix] (comparison) 
				{Group Comparisons: HC vs MCI vs AD, Motor Execution vs Motor Imagery};
				\node[detail, fill=green!5, below=0.3 of comparison] (visualization) 
				{Visualization: Circular graphs, Edge difference matrices};
				
				\draw[arrow, blue, opacity=0.5] (acquisition.south) -- (devices.north);
				\draw[arrow, blue, opacity=0.5] (devices.south) -- (sync.north);
				\draw[arrow, orange, opacity=0.9] (sync.south) -- (preprocessing.north);
				\draw[arrow, orange, opacity=0.9] (preprocessing.south) -- (source.north);
				\draw[arrow, orange, opacity=0.9] (source.south) -- (avalanche.north);
				\draw[arrow, orange, opacity=0.7] (avalanche.south) -- (criticality.north);
				\draw[arrow, green!90!black] (criticality.south) -- (atm.north);
				\draw[arrow, green!90!black] (atm.south) -- (matrix.north);
				\draw[arrow, green!90!black] (matrix.south) -- (comparison.north);
				\draw[arrow, green!90!black] (comparison.south) -- (visualization.north);
				
			\end{tikzpicture}
			\caption{Analytical pipeline for multimodal neuronal avalanche analysis: from synchronized EEG-fNIRS acquisition to network construction in Alzheimer's disease.}
			\label{fig:analytical_pipeline}
		\end{figure}

		An integrated EEG-fNIRS framework is implemented here (Figure~\ref{fig:analytical_pipeline}) to analyze neuronal avalanche dynamics and assess large-scale network function. The methodological approach combines: (1) precise synchronization of EEG and fNIRS to capture multimodal neural dynamics, (2) application of avalanche analysis to both electrophysiological and hemodynamic signals, and (3) examination of motor network flexibility through execution-imagery contrasts as a potential marker of early network disruption. This integrated framework aims to detect functional network alterations that may precede structural changes in the AD continuum.
		
		\subsection{A Priori Sample Size Estimation}
		To ensure scientific rigor and adequate statistical power for the planned validation study, an a priori power analysis was conducted using an online-available resting-state EEG dataset of AD patients and healthy controls (HC) \cite{b8}. The analysis employed the theta-to-alpha ratio (TAR), a well-established spectral biomarker of AD-related cognitive decline characterized by oscillatory slowing. Monte Carlo simulations using fitted TAR distributions indicated that 60 participants (30 HC, 30 AD) provide 80\% power at $\alpha = 0.05$ to detect expected group differences \cite{b9}. The MCI group was set to 30 participants to maintain balanced design for planned comparisons.
		
		\subsection{Participants and Experimental Paradigm}
		\subsubsection{Pilot Cohort}
		This proof-of-concept study reports data from a pilot cohort of four participants: one with mild cognitive impairment (MCI), one neurologically intact patient, and two healthy controls. This small cohort enables the testing and refinement of the synchronized EEG-fNIRS acquisition and multimodal avalanche analysis pipeline. Observations from this group are therefore presented as illustrative case examples to inform the design of future, statistically powered studies.
		
		\subsubsection{Motor Task Design}
		Participants performed a custom-designed motor task involving both execution and imagery of right/left hand grasping movements within an engaging visual feedback environment. Visual targets (``lanterns") prompted participants to control a cursor (``rabbit") through either actual or imagined hand movements. This BCI-based paradigm robustly engages sensorimotor networks and generates well-characterized event-related desynchronization in mu/beta bands, providing a reliable foundation for assessing network dynamics across movement conditions.
		
		\subsection{Multimodal Data Acquisition and Synchronization}
		Neural data were acquired using synchronized EEG and fNIRS systems. EEG was recorded using a 64-channel Brain Products actiCHamp system with electrodes providing global head coverage according to the international 10-10 system. Concurrent fNIRS data were acquired using a NIRx Aurora system with 16 sources and 16 detectors arranged in a customized montage over bilateral sensorimotor cortex areas. The fNIRS optode placement followed an extended 10-10 system nomenclature, ensuring comprehensive coverage of motor-related regions. Precise temporal synchronization was achieved through a digital trigger protocol cycling markers from 1-255, creating unambiguous temporal alignment between systems with $\sim$60 ms accuracy \cite{b10}. 
		
		\subsection{Signal Pre-Processing Pipeline}
		EEG signals were band-pass filtered (8-30 Hz) to focus on sensorimotor rhythms relevant to motor processing. Cortical source activity was reconstructed using an average template head model, with the cortex parcellated into regions of interest according to the Desikan-Killiany atlas \cite{b11}. This represents a standard methodological approach, though the use of a template rather than individual anatomy is noted as a limitation, particularly for elderly and MCI/AD populations where anatomical variability may be greater. Regional time series were extracted as the principal component of dipoles within each region of interest (ROI).
		
		The fNIRS processing pipeline converted raw light intensity to optical density, applied motion artifact correction using temporal derivative distribution repair, and derived hemoglobin concentrations (HbO, HbR) via the modified Beer-Lambert law. Bandpass filtering (0.01-0.2 Hz) preserved hemodynamic responses while removing drift and high-frequency noise. Continuous data were segmented into epochs time-locked to motor execution and imagery events.
		
		\subsection{Neuronal Avalanches }
		
		\subsubsection{Theoretical Framework}
		The analysis of large-scale brain dynamics builds upon the framework of neuronal avalanches, which characterizes spontaneous and task-evoked neural activity as cascades of activations propagating through brain networks. This approach is grounded in evidence that the brain operates near a critical state optimal for information processing, and that neurodegenerative diseases may disrupt this critical balance \cite{b12, b13, b14}.
		
		\subsubsection{Avalanche Detection}
		For both EEG and fNIRS modalities, regional time series were z-scored and binarized using a pre-defined threshold, typically set to \( |z| > 3 \). Avalanches -- sequences of consecutive time bins containing activity bounded by silent periods -- were identified to test the hypothesis of brain operation near a critical point. This dynamical regime, where activity is balanced at an ``edge of instability", is characterized by scale-invariant propagation patterns observed across neural recordings \cite{b14, b15}. Temporal scales were set according to signal physiology: millisecond bins captured neuro-electrical propagation for EEG, while second-scale bins captured the slower hemodynamic waves for fNIRS \cite{b16}.
		
		\subsubsection{Quantifying Propagation Dynamics}
		Propagation dynamics were quantified using the branching parameter $\sigma$. For a single avalanche $i$, the branching parameter is calculated as the geometric mean of the ratio of activations between subsequent time bins \cite{b17, b18}:
		
		\begin{equation}
			\sigma_i = \left( \prod_{j=1}^{N_{\text{bin}}^{(i)}-1} \frac{n_{\text{events}}^{(i)}(j+1)}{n_{\text{events}}^{(i)}(j)} \right)^{\frac{1}{N_{\text{bin}}^{(i)}-1}}
			\label{eq:sigma_i}
		\end{equation}
		
		The global branching parameter $\sigma$ for a participant's recording is then the geometric average over all $N_{\text{avail}}$ avalanches \cite{b19}:
		
		\begin{equation}
			\sigma = \left( \prod_{i=1}^{N_{\text{avail}}} \sigma_i \right)^{\frac{1}{N_{\text{avail}}}}
			\label{eq:sigma_global}
		\end{equation}
		
		\noindent where $N_{\text{bin}}^{(i)}$ is the total number of time bins in the $i$-th avalanche, and $n_{\text{events}}^{(i)}(j)$ is the number of active regions in the $j$-th bin of that avalanche. A branching parameter $\sigma \approx 1$ indicates that the neural activity is propagating in a critical state.
		
		\subsection{Avalanche Transition Matrix Construction}
		ATMs were constructed to quantify transition probabilities between brain regions: $P(j\ \text{active at}\ t+\delta \mid i\ \text{active at}\ t)$. This creates directed, weighted networks representing information flow pathways. The matrices provide sensitivity to dynamic propagation patterns compared to traditional static connectivity measures. Critically, this framework is agnostic to the signal origin: ATMs were constructed from regional time series derived from either the preprocessed fNIRS hemoglobin concentrations (oxygenated HbO and deoxygenated hemoglobin HbR) or the source-localized EEG activity. Separate matrices were computed for the motor execution and motor imagery conditions, providing the basis for the subsequent contrast analysis of network dynamics between tasks.

		\section{Results}
		
		The primary aim of this pilot analysis is to demonstrate that the BCI paradigm can yield coherent, multimodal neural measures. The results shown below, while from a limited sample, establish this foundational feasibility and provide a clear basis for expanded investigation.
		
		\subsection{Multimodal Convergence in Network Dynamics}
		Network dynamics were assessed by computing the condition contrast (Motor Imagery minus Motor Execution). This analysis revealed a convergent pattern across fNIRS and EEG data in the available participants: the neural distinction between motor execution and imagery appeared diminished in the MCI participant compared to the healthy control (HC) participant for whom full data is presented.
			
\begin{figure}[htbp]
			\centerline{\includegraphics[width=9cm]{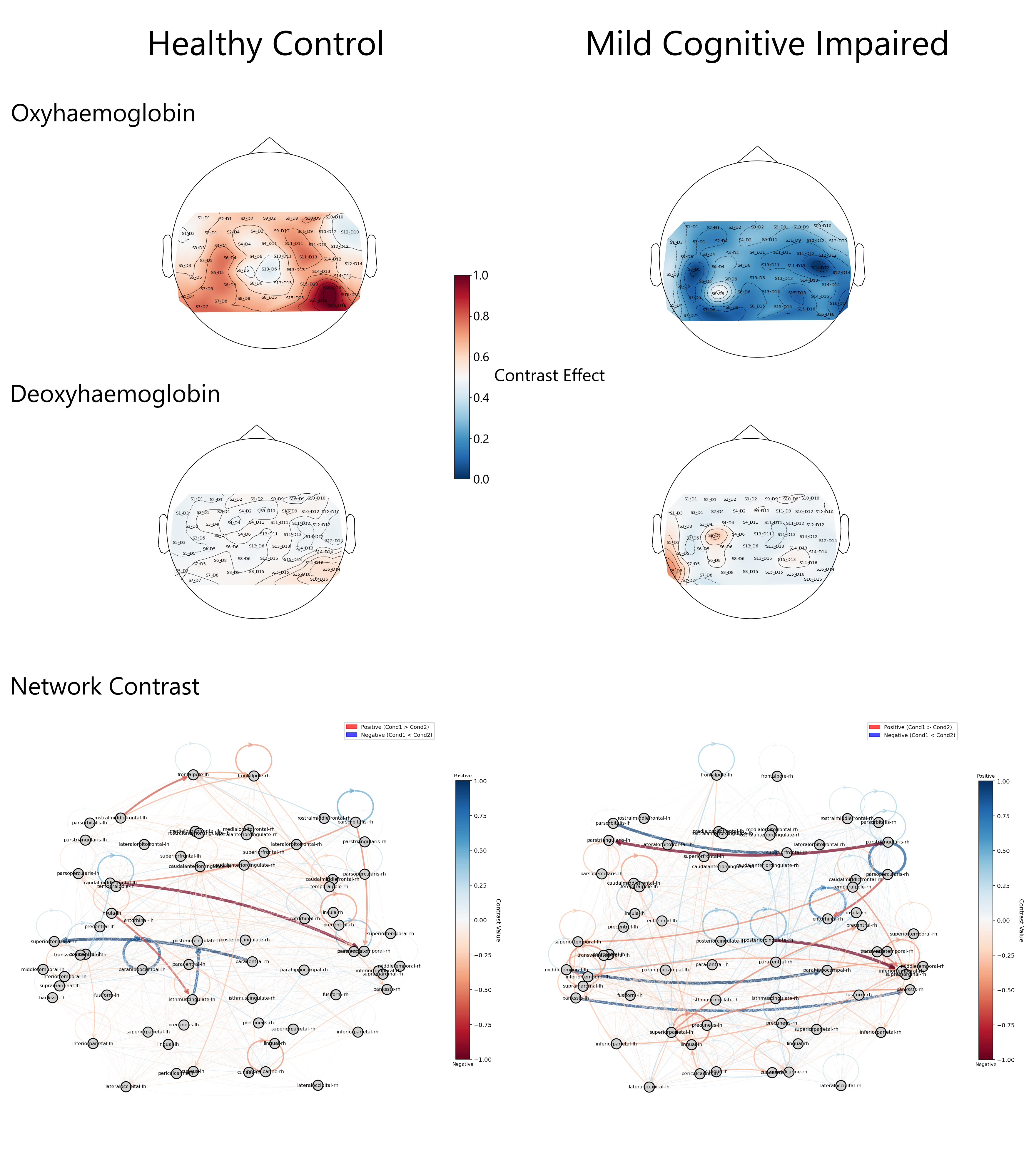}}
			\caption{fNIRS and EEG condition contrasts (Motor Imagery – Execution). Contrast values for HbO (top) and HbR (middle), and ATM-derived source-localized EEG connectivity maps (bottom) are shown for a representative healthy control (left) and mild cognitive impaired participant (right)}
			\label{fig:multimodal_contrast}
		\end{figure}

		For fNIRS, the distribution of oxyhemoglobin (HbO) contrast values peaked around 1 for the HC, indicating a consistent directional difference between conditions. For the MCI participant, the distribution centered around 0, suggesting a less pronounced differentiation (Figure ~\ref{fig:multimodal_contrast}).
		
		A convergent electrophysiological pattern was observed. The network contrast map, derived from ATM analysis of source-localized EEG, showed more distinct regional differentiation in the healthy controls (as illustrated for one HC, with similar trends in others). In contrast, the map for the single MCI participant demonstrated more spatially common activity. This persistent, condition-specific alteration across both fNIRS and EEG modalities indicates a potential disruption in the motor system's ability to reconfigure between cognitive states, a pattern that merits investigation in larger cohorts \cite{b20, b21}.
		
		\subsection{Single-Trial Functional Connectivity}
		
		Analysis of single-trial functional connectivity provides insight into the consistency of neural network organization during motor tasks. The Phase Lag Index was employed to examine communication patterns between brain regions across individual trials, revealing the dynamic characteristics of neural coordination during repeated task performances \cite{b22}.
		
\begin{figure}[htbp]
			\centerline{\includegraphics[width=9cm]{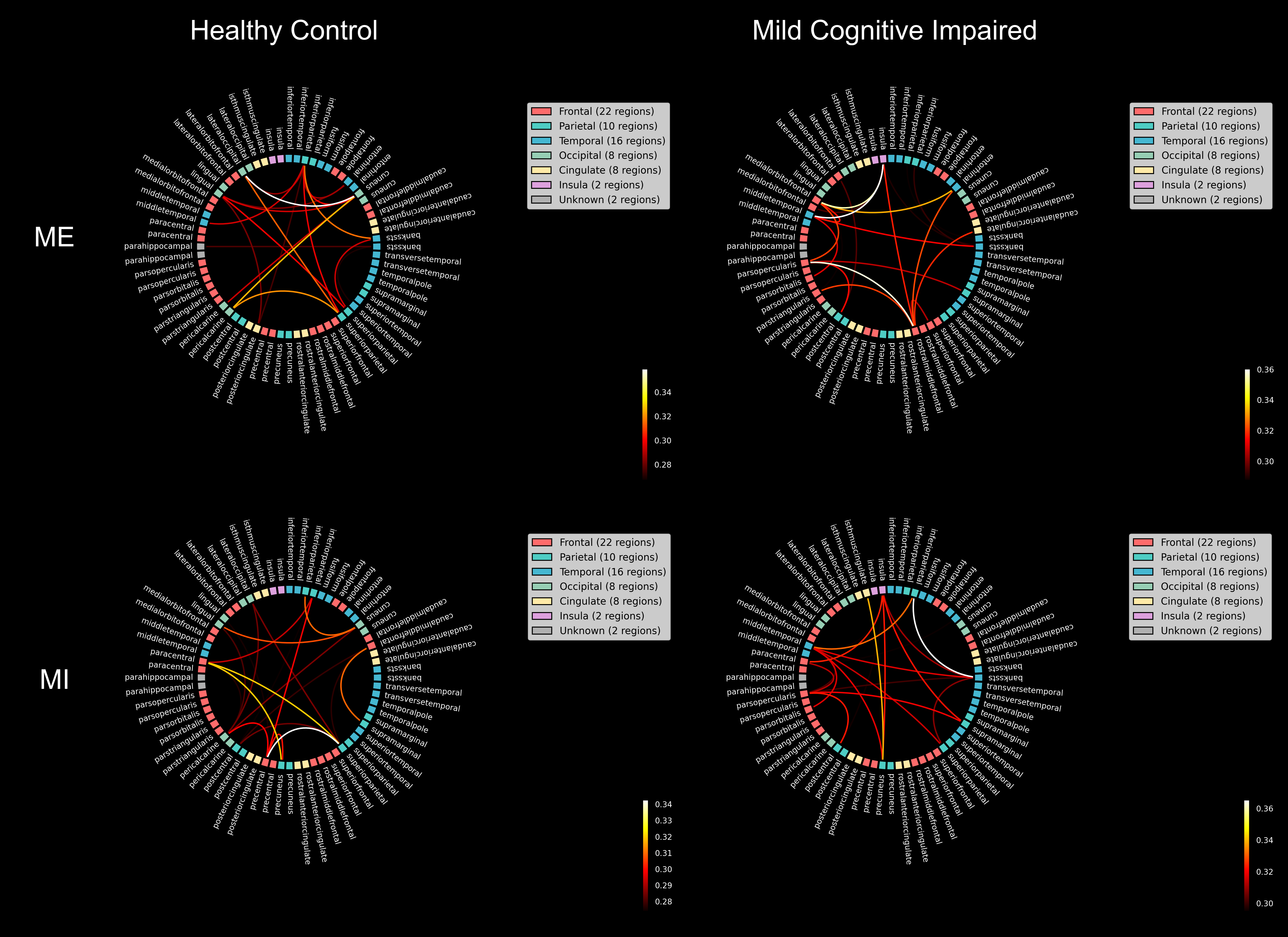}}
			\caption{Single-trial network connectivity across motor conditions. (A) Healthy control demonstrates consistent Phase Lag Index connectivity patterns across individual trials of both execution and imagery tasks, indicating robust neural dynamics. (B) Mild cognitive impairment participant exhibits substantial trial-to-trial variability and condition-dependent network reorganization, particularly during motor imagery, suggesting unstable neural computation in early cognitive impairment.}
			\label{fig:network_plots}
		\end{figure}

		As illustrated in Figure~\ref{fig:network_plots}, distinct patterns emerge between the two participants. The healthy control demonstrates consistent connectivity profiles across motor execution and imagery conditions, with stable engagement of frontoparietal networks. This consistency suggests maintained neural coordination during both physical and imagined motor acts.
		
		In comparison, the participant with mild cognitive impairment exhibits substantial trial-to-trial variability, particularly during motor imagery. The network organization shows increased variation from the execution condition, suggesting potential differences in neural recruitment during cognitive motor tasks. This pattern suggests that, in this illustrative case, the reduction in mean differentiation between task states was accompanied by increased dynamic instability from trial to trial, presenting a dual signature of altered network dynamics.

		\section{Discussion}
		
		This proof-of-concept study demonstrates that combining high-density EEG-fNIRS with motor paradigms offers a promising approach for detecting early network dysfunction in Alzheimer's disease. While memory and language tasks have traditionally dominated AD research, motor execution and imagery provide distinct advantages as sensitive probes of neural integrity. These paradigms engage frontoparietal networks known to be affected in early AD while minimizing confounding factors like educational background and cultural influences that often complicate cognitive assessment. More importantly, motor imagery specifically requires intact predictive coding and internal model maintenance -- precisely the cognitive processes that may be most vulnerable in early neurodegeneration.
		
		The rationale for employing synchronized high-density EEG and fNIRS recordings stems from their complementary capabilities to capture both neural electrical activity and hemodynamic responses within the same experimental session. The comprehensive spatial coverage afforded by 64 EEG electrodes and 16 fNIRS source-detector pairs enables detection of fine-grained network patterns that might be obscured in sparser configurations. This high resolution is particularly crucial for resolving subtle alterations that could otherwise be masked by inter-subject anatomical variability or measurement noise. Furthermore, the simultaneous acquisition provides a unique window into neurovascular coupling -- the fundamental relationship between neural activity and subsequent hemodynamic responses -- which has been increasingly implicated in neurodegenerative processes \cite{b23}.
		
		The experimental design also incorporated EMG recordings, which, while not analyzed in this preliminary study, establish a foundation for future investigations linking neural dynamics to behavioral outputs. The potential to extract precise reaction times and movement kinematics from these recordings could provide crucial behavioral correlates to complement the neural measures, creating a comprehensive framework that spans from network-level processing to motor execution.
		
		An integrated interpretation of the pilot data yields a convergent picture. The dual observations -- a reduced average neural contrast between motor states coupled with increased trial-to-trial variability in the single MCI case -- are likely complementary facets of declining network health. This pattern, where task states are less distinct on average and  exhibit greater trial-to-trial variability, aligns with a degradation of dynamic flexibility and could represent a functional correlate of the early cognitive ``disconnection syndrome." The observed reduction in network contrast invites several non-exclusive mechanistic hypotheses for future testing: network dedifferentiation (a loss of functional specialization), compensatory recruitment during demanding imagery, or a predictive coding deficit impairing internal model generation.
			
		The findings and their interpretation are inherently shaped by the proof-of-concept design. The primary constraint is the small pilot sample size, which, while enabling technical validation, prevents broad generalization. This makes it challenging to distinguish potential mechanistic signals from individual variability at this stage. Consequently, future research must prioritize validation in a larger, statistically powered cohort to test whether the observed patterns are replicable biomarkers. The planned study with 90 participants (30 per group) will address this core need. Subsequent work will also aim to simplify the hardware configuration for practicality and explore integrating behavioral measures to strengthen the clinical relevance of the neural metrics.

	\end{document}